\documentclass[12pt,a4paper]{article}
\usepackage{graphicx}
\usepackage{authblk}
\usepackage{amsfonts}
\usepackage[dvips]{color}

\def\ket#1{|\,#1\,\rangle}
\def\bra#1{\langle\, #1\,|}

\def\proj#1#2{\ket{#1}\bra{#2}}

\newcommand{\beq}{\begin{equation}}
\newcommand{\eeq}{\end{equation}}
\newcommand{\beqa}{\begin{eqnarray}}
\newcommand{\eeqa}{\end{eqnarray}}

\begin{document}

\title{Determining the output port from the distance}
\author[]{Iulia Ghiu}
\affil[]{{University of Bucharest, Faculty of Physics, PO Box MG-11, R-077125, Magurele, Romania }\\
Email: {\em iulia.ghiu@g.unibuc.ro}}

\maketitle
\begin{abstract}
{\it Abstract.} In this paper we discuss the following scenario. Suppose that two distant observers share an entangled state of two $D$-level systems. The two observers, Alice and Bob, have to send randomly their particles through a multi-input multi-output device. The task is to determine the nature of this device or black-box, such that Alice will be able to determine with certainty the output port of Bob's particle. We prove that this black-box is the multi-input-port quantum sorter. Further, we investigate if it is possible to determine the output port in the case when three observers are involved by analyzing the cases when the Greenberger-Horne-Zeilinger (GHZ) state and the W state are used.
\end{abstract}

\section{Introduction}

Multipartite entanglement plays a central role in quantum information processing \cite{Nielsenbook}. Determining the set of the mutually unbiased bases is of great importance in quantum cryptography or tomography \cite{Ghiu-2012}, \cite{Ghiu-2013}, \cite{Ghiu-math-2014}, \cite{Ghiu-phys-2014}, \cite{Manko-1-2014}, \cite{Manko-2-2014}. The multiparticle continuous and discrete variable quantum systems also find applications in quantum information theory and therefore quantifying these resources became imperative. Non-Gaussianity \cite{Marian-2014}, entanglement \cite{Isar-2015}, \cite{Ghiu-2015}, \cite{Isar-rom-j-2019}, quantum discord \cite{Isar-eur-2017}, \cite{Serban-2015}, \cite{Messina-2019}, steerability \cite{Isar-eur-2018}, \cite{Isar-rom-j-2018}, quantum degree of polarization \cite{Ghiu-2016}, \cite{Ghiu-pra-2018}, \cite{Ghiu-rom-rep-2018}, interferometric power \cite{Isar-rom-rep-2018}  are only a few examples of measures recently investigated.

\begin{figure}[thb]
\centering
\includegraphics[scale=0.5]{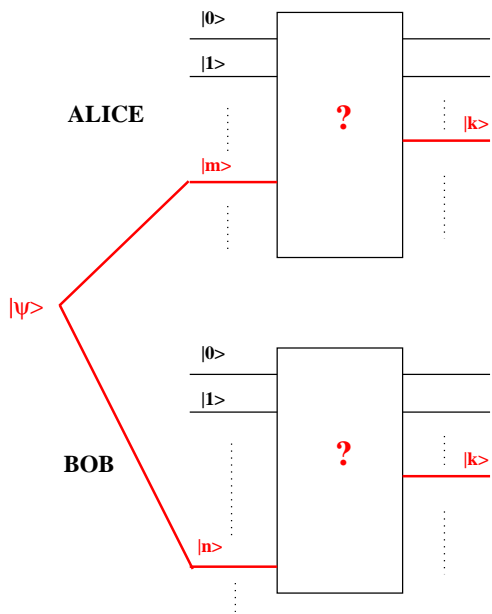}
\caption{Illustration of the setup raised by the question given in the Introduction. Alice and Bob share an entangled two-quDit state $\ket{\psi}_{AB}$. Each of the particles is randomly injected on a multi-input multi-output black-box. If Alice' particle exits through the mode $\ket{k}$, then she knows with certainty that Bob's particle also emerges through the output $\ket{k}$. The problem is to determine the nature of this black-box.}
\label{fig-introd}
\end{figure}

In this paper, we want to answer the following question.
Suppose that two observers Alice and Bob, situated at different locations, have a black-box, which is a multi-input multi-output device. Alice and Bob hold one particle of the two-quDit system (D-level quantum system) found in the entangled state:
\[
\ket{\psi }_{AB}= \sum_{j=0}^{D-1} \alpha _j \, \ket{j}_A\ket{j}_B,
\]
with $\sum_j|\alpha_j |^2=1$. Alice and Bob send their particle to their black-box, by randomly choosing the input port. Suppose that $\ket{m}$ is the input of Alice' particle and $\ket{n}$ the input of Bob's particle. How should the black-box be constructed such that the two particles will exit through the same output, denoted by $\ket{k}$, in the two different laboratories? Or, in other words, if Alice' particle emerges through the mode $\ket{k}$, then she knows with certainty that the output of Bob's particle is also $\ket{k}$, as it is shown in Fig. \ref{fig-introd}.

We show that the answer to the above question is given by the fact that the black-box of Fig. \ref{fig-introd} should be replaced by the multi-input-port quantum sorter of quDits, which was defined by us in Ref. \cite{Ghiu-sorter-2019}. The quantum sorter is a device, which selects the states in different output ports according to their properties.

Different kinds of quantum sorter were devoted to the study of the properties of the photons: polarization, orbital angular momentum (OAM) \cite{Leach-2002}, \cite{Berkhout}, \cite{Lavery}, \cite{Boyd-nat-com-2013}, \cite{Schulz}, \cite{Wang}, the total angular momentum \cite{Leach-2004}, or radial modes \cite{Zhou}, \cite{Gu-2017}, \cite{Boyd-2019}.
In addition, a more general quantum sorter was defined in Ref. \cite{Ionicioiu}, which performs the selection of the states for any degree of freedom of a single quDit. This sorter works properly only in the case when the particles are incident on a single input port.

The paper is organized as follows. Section 2 presents a brief review of quantum sorter. In addition, we show that the quantum circuit of the multi-input-port quantum sorter consists of two controlled gates. In Section 3, it is proved that the multi-input-port quantum sorter is the proper device that accomplishes the task raised in the Introduction. Namely, after Alice and Bob send randomly their particles through any of the input-ports of the quantum sorter, the two particles will exit through the same output-port. This allows Alice to determine the output port of Bob's particle from the distance. Further, Section 4 is devoted to the study of the case when three parties are involved. When the GHZ state is used by the three observers, Alice can infer the output port of the other two distant observers.
Our conclusions are outlined in Sec. 5.

\section{Quantum sorter}

Consider a $D$-level quantum system, whose states are described by the computational basis $\{ \ket{0}$, $\ket{1}$,..., $\ket{D-1}\}$ in the Hilbert space ${\cal H}_{system}$. In addition, let us denote the states of $D$ ports by $\ket{j}$, with $j$= 0, 1, ..., $D-1$ acting in the Hilbert space ${\cal H}_{port}$.
The quantum sorter has the property to select the particles characterized by the property $\ket{s}$ of the quantum system by collecting them to the same output port-state denoted by $\ket{s}$.

Consider two $D$-dimensional Hilbert spaces ${\cal H}_{system}$ and ${\cal H}_{port}$, the first one associated with a quDit system, while the second one to the states of the $D$ ports.
A device acting in the Hilbert space ${\cal H}_{system}\otimes {\cal H}_{port}$ described by the unitary operator $U$:
\beq
U\, \ket{s}\ket{*}=\ket{\#}\ket{s}, \; \; \; \mbox{with} \; s=0,1,..., D-1
\label{definitia-sorter}
\eeq
is called {\bf quantum sorter} \cite{Ghiu-sorter-2019}. The symbols $*$ and $\#$ are allowed to take the values 0, 1, 2, ..., $D-1$. In the particular case when $\ket{*}=\ket{\#}=\ket{n}$, the SWAP gate is obtained:
\[
\mbox{SWAP}\,  \ket{s}\ket{n}=\ket{n}\ket{s}.
\]

The perfect multi-input-port quantum sorter is a device that
sends the particle found in the state $\ket{s}$, through the output port-state $\ket{s}$, regardless of the input port-state, such that the state of the particle remains unchanged:

\[
\ket{s}\ket{k} \longrightarrow \ket{s}\ket{s}, \; \; \; \mbox{with} \; s, k=0,1,..., D-1
\]
In Ref. \cite{Ghiu-sorter-2019}, it was shown that the perfect quantum sorter cannot be constructed.

A device acting on the Hilbert space ${\cal H}_{system}\otimes {\cal H}_{port}$
that performs the following unitary transformation:
\beq
U_{SQS}\, \ket{s}\ket{k} = \ket{s}\ket{s \oplus k}= \left\{ \begin{array}{lcl}
\ket{s}\ket{s}, & \mbox{if} & k=0 \\
\ket{s}\ket{j}, &\mbox{with} \; j=s \oplus k\ne s, \; \mbox{if} & k\ne 0,
\end{array} \right.
\label{single-sorter}
\eeq
with $k=0$ and $s$ = 0, 1,..., $D-1$, is called {\bf the single-input-port quantum sorter (SQS)} \cite{Ghiu-sorter-2019}, \cite{Ionicioiu}. The incident particles must be incident only on the input port-state $\ket{0}$.

The controlled-$U$ gate acting on two quDits is defined as \cite{Nielsenbook}, \cite{Ghiu-sorter-2019}:
\[
C(U)\, \ket{s}\otimes \ket{k}=\ket{s}\otimes \, U^s \ket{k},
\]
where $\ket{s}$ is the control quDit, while $\ket{k}$ is the target quDit. Let us denote the controlled-$U$ gate acting on the first quDit, which is the target and on the second quDit, which is the control, as follows:
\[
\tilde C(U)\, \ket{s}\otimes \ket{k}=U^k\ket{s}\, \otimes \ket{k}.
\]

 In Ref. \cite{Ionicioiu}, it was proved that $U_{SQS}=C(X_D)$, where $X_D$ is the generalized Pauli operator, i.e. $X_D=\sum_{j=0}^{D-1}\, \proj{j\oplus 1}{j}$.

A device acting in the Hilbert space ${\cal H}_{system}\otimes {\cal H}_{port}$, that performs the unitary transformation
\beq
U_{MQS}\, \ket{s}\ket{k} = \ket{s \ominus k}\ket{s}, \; \; \; \mbox{with} \; s, k=0,1,..., D-1
\label{m-sorter}
\eeq
is called {\bf the multi-input-port quantum sorter (MQS)} \cite{Ghiu-sorter-2019}.
This quantum sorter sends the particle found in the state $\ket{s}$, through the output state $\ket{s}$, regardless of the input port state, while the state of the particle is modified.

We have proved in Ref. \cite{Ghiu-sorter-2019} a theorem, that shows how the multi-input-port quantum sorter can be constructed with the help of quantum gates, namely:
\beq
U_{MQS}\ket{s}\ket{k}=U_{SQS}\, \left( (X_D^\dagger )^k\otimes I\right)\,\ket{s}\ket{k}.
\label{th}
\eeq

\begin{figure}[thb]
\centering
\includegraphics[scale=0.6]{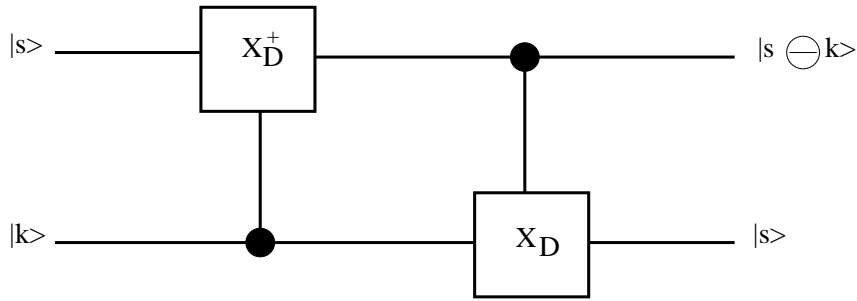}
\caption{The quantum circuit of the multi-input-port quantum sorter based on two controlled gates, according to Eq. (\ref{expr-gen-multi-sorter}). }
\label{fig-multi-sorter}
\end{figure}

According to Eq. (\ref{th}), the expression of the multi-input-port quantum sorter is given by the unitary operator:
\beq
U_{MQS}=C(X_D)\, \tilde C(X_D^\dagger ).
\label{expr-gen-multi-sorter}
\eeq
In Fig. \ref{fig-multi-sorter} it is presented the quantum circuit of the multi-input-port quantum sorter.

\section{Determining the output port from the distance}

In this section, we answer the question raised in the Introduction. Consider that there is entanglement between two quDits shared by Alice and Bob, situated at different locations, the state of the whole system being:
\beq
\ket{\psi }_{AB}= \sum_{j=0}^{D-1} \alpha _j \, \ket{j}_A\ket{j}_B,
\label{ent}
\eeq
with $\sum_j|\alpha_j |^2=1$.
For example, this state can be generated by using OAM of two photons \cite{Zeilinger-light-2018}.

Let us investigate the following scenario: Alice and Bob have a $D$-input $D$-output black-box in their own lab. They are allowed to send their particle of the entangled state (\ref{ent}) randomly through any input port. Alice has to guess the output port of the second particle situated in Bob's laboratory. What is the nature of the black-box, such that Alice guesses the correct output port of Bob's particle with certainty?

\begin{figure}[thb]
\centering
\includegraphics[scale=0.5]{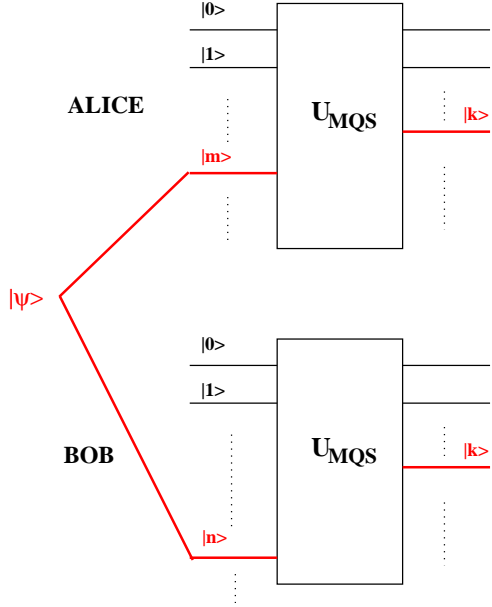}
\caption{The scheme for determining the output port by a distant observer. The black-box in Fig. \ref{fig-introd} is represented by the multi-input-port quantum sorter of quDits $U_{MQS}$ shown in Fig. \ref{fig-multi-sorter}. After Alice' particle exits through the output $\ket{k}$, then she knows with certainty that the second particle situated in Bob's lab also emerges through the output $\ket{k}$. }
\label{fig-ent-sorter}
\end{figure}

Each Alice and Bob constructs the multi-input-port quantum sorter of quDits of Eq. (\ref{expr-gen-multi-sorter}) in their lab, according to the scheme presented in Fig. \ref{fig-multi-sorter}. Alice and Bob send their particles on the universal multi-input-port quantum sorter by randomly choosing the input port. Let us denote by $\ket{m}$ the input of Alice' particle and by $\ket{n}$ the input of Bob's particle, as it is shown in Fig. \ref{fig-ent-sorter}.

Let us evaluate the final state after applying the two quantum sorters $U_{MQS}^{(A)}$ and $U_{MQS}^{(B)}$.
According to Eq. (\ref{m-sorter}), one obtains:
\[
U_{MQS}^{(A)}\otimes U_{MQS}^{(B)}\ket{\psi }_{AB}\ket{m}_A\ket{n}_B=\sum_{j=0}^{D-1} \alpha _j \, \ket{j\ominus m}_A\ket{j}_A\otimes \ket{j\ominus n}_B\ket{j}_B.
\]

Thus, if Alice gets a click in the detector situated in the output $\ket{k}$, then the final state reads:
$$\ket{k\ominus m}_A\ket{k}_A\otimes \ket{k\ominus n}_B\ket{k}_B. $$
The probability that Alice would detect a particle in the detector $\ket{k}$ is $|\alpha_k|^2$, this $k$ being arbitrary, any of 0, 1,..., $D-1$. Therefore, regardless of the value of $k$, Alice will know with certainty that Bob will get a click also in the detector $\ket{k}$, as one can see in Fig. \ref{fig-ent-sorter}. This scheme allows the determination of the output port, with the probability equals unity, by a distant observer.

\section{Generalization to the case of three-partite systems}

In the case of three qubits, it was shown that there are two inequivalent classes of entangled states: the GHZ class and the W class \cite{Vidal}. The Greenberger-Horne-Zeilinger (GHZ) state is defined as:
\beq
\ket{GHZ}=\frac{1}{\sqrt 2}\, (\ket{000}+\ket{111}),
\label{st-ghz}
\eeq
while the W state is given by
\beq
\ket{W}={1\over \sqrt{3}}\, (\ket{001}+\ket{010}+\ket{100}).
\label{st-w}
\eeq
The concept of 'inequivalent' means that if we allow only stochastic local quantum operations and classical communications, then one cannot transform states from the GHZ-class to the W-class and vice-versa with a non-zero probability of success \cite{Vidal}. In addition, in Ref. \cite{Ghiu-2001} it was studied the possibility of transforming between the two inequivalent classes using a catalysis state.

\subsection{The GHZ state}

Suppose that a three-particle system found in the GHZ state of Eq. (\ref{st-ghz}) is shared by three distant observers denoted by Alice, Bob, and Charlie. Further, each of them has a multi-input-port quantum sorter for qubits. The state after the three particles enter the devices is:
\beqa
&&U_{MQS}^{(A)}\otimes U_{MQS}^{(B)}\otimes U_{MQS}^{(C)}\ket{GHZ }_{ABC}\ket{m}_A\ket{n}_B\ket{p}_C \nonumber \\
&=&\frac{1}{\sqrt 2}\, \left( \ket{m}\ket{0}\right)_A\left( \ket{n}\ket{0}\right)_B\left( \ket{p}\ket{0}\right)_C \nonumber \\
&&+ \frac{1}{\sqrt 2}\, \left( \ket{1\ominus m}\ket{1}\right)_A\left( \ket{1\ominus n}\ket{1}\right)_B\left( \ket{1\ominus p}\ket{1}\right)_C. \label{fin-ghz}
\eeqa
We have denoted by $m$ the input port of Alice' particle, by $n$ the input port of Bob's particle and $p$ the input port of Charlie's particle.

With the probability 1/2, Alice will get a click in the detector $\ket{0}$, this leading to the final state
$$\left( \ket{m}\ket{0}\right)_A\left( \ket{n}\ket{0}\right)_B\left( \ket{p}\ket{0}\right)_C .$$
As one can see, both Bob and Charlie will detect their particles in the same output port as Alice, i.e. $\ket{0}$.

Also, Alice detects the particle in the output port $\ket{1}$ with the probability 1/2. In addition, Bob and Charlie get a click in the output port $\ket{1}$ of their device. This means that Alice can infer the output port of the two distant observers Bob and Charlie. We need to emphasize the the states of the three qubits are modified after passing through the multi-input-port quantum sorters.

\subsection{The W state}

Let us invetigate a similar scenario, where in this case the initial three-qubit state is the W state of Eq. (\ref{st-w}).

The final state after passing through the multi-input-port quantum sorters is given by:
\beqa
&&U_{MQS}^{(A)}\otimes U_{MQS}^{(B)}\otimes U_{MQS}^{(C)}\ket{W}_{ABC}\ket{m}_A\ket{n}_B\ket{p}_C \nonumber \\
&=&\frac{1}{\sqrt 3}\, \left( \ket{m}\ket{0}\right)_A\left( \ket{n}\ket{0}\right)_B\left( \ket{1\ominus p}\ket{1}\right)_C \nonumber \\
&&+ \frac{1}{\sqrt 3}\, \left( \ket{m}\ket{0}\right)_A\left( \ket{1\ominus n}\ket{1}\right)_B\left( \ket{p}\ket{0}\right)_C\nonumber \\
&&+ \frac{1}{\sqrt 3}\, \left( \ket{1\ominus m}\ket{1}\right)_A\left( \ket{n}\ket{0}\right)_B\left( \ket{p}\ket{0}\right)_C. \label{fin-w}
\eeqa

With the probability 1/3, Alice and Bob will get a click in the detector $\ket{0}$, while Charlie a click in $\ket{1}$. The output port of two observers cannot be determined only by the other (third) observer. Therefore, the scheme which uses the three-particle W state fails in getting the right answer regarding the output ports of distant observers.

\section{Conclusions}

We have investigated the possibility of determining the output port of a particle by a distant observer. Suppose that two separated observers share a two-particle system of quDits found in an entangled state. Each of the two observers holds a black-box, which is a multi-input multi-output device. The two observers randomly send their particle through any input port of the device. How this black-box has to be constructed, such that one of two observers will determine with certainty the output port of the particle of the other distant observer?

We have shown in this paper that the black-box has to be the multi-input-port quantum sorter. At the end of the protocol, the states of the two particles are modified.

Further, we have investigated the case of three-partite systems, where the states of the particles belong to the two inequivalent classes of three qubits: the GHZ class and the W class. Suppose that the three particles are shared by the distant observers: Alice, Bob, and Charlie. We have proved that only for the GHZ state, Alice can detect with certainty the output port of the particles of both Bob and Charlie by using also the multi-input-port quantum sorter. For the W state the protocol fails.

\end{document}